\documentclass[aps, prb, superscriptaddress, twocolumn, letterpaper, 10pt, amsfonts, amsmath, amssymb]{revtex4-2}
\usepackage{graphicx}
\usepackage[colorlinks=true,citecolor=blue,urlcolor=blue,linkcolor=blue]{hyperref}
\usepackage{physics}
\usepackage{float}
\usepackage{xcolor}
\usepackage{soul}

\definecolor{ABcolor}{cmyk}{1,0,0,0.6}

\newcommand{\ii}{\mathrm{i}}

\renewcommand{\eqref}[1]{{Eq.~(\ref{#1})}}
\newcommand{\figref}{Fig.~\ref}

\begin{document}
\title{Spontaneous edge corner currents in $s+is$ superconductors and time reversal symmetry breaking surface states}

\author{Andrea Benfenati} \email{alben@kth.se}
\affiliation{Department of Physics, The Royal Institute of Technology, Stockholm SE-10691, Sweden}
\author {Egor Babaev}
\affiliation{Department of Physics, The Royal Institute of Technology, Stockholm SE-10691, Sweden}

\begin{abstract}
We present a study of the basic microscopic  model of an $s$-wave  superconductor with frustrated interband interaction.
When frustration is strong, such an interaction gives raise to a $s+is$ state.
This is a $s$-wave superconductor that spontaneously breaks time reversal symmetry. We show that, in addition to the known
$s+is$ state, there is additional phase, where 
the system's bulk is a conventional $s$-wave state, but superconducting surface states break time reversal symmetry. Furthermore, we show that $s+is$ superconductors can have spontaneous boundary currents and spontaneous magnetic fields. These arise at lower-dimensional boundaries, namely, the corners, in two-dimensional samples. This demonstrates that boundary currents  effects in superconductors can arise  in states which are not topological and not chiral according to the modern classification.
\end{abstract}

\maketitle
\section{Introduction}
The most common type of superconductivity occurs where electrons form spin-singlet Cooper pairs.
Such a superconductor spontaneously breaks local $U(1)$ symmetry. 
Recent experiments reported the discovery of the so-called $s+is$ superconductor  Ba$_{1-x}$K$_x$Fe$_2$As$_2$ \cite{grinenko2020superconductivity,grinenko2017superconductivity,grinenko2021bosonic}.
The $s+is$  superconductor \cite{StanevTesanovic,carlstrom2011length, maiti2013s+,boker2017s+} is a spin singlet superconductor, that  spontaneously breaks an additional time reversal symmetry, so the total broken symmetry becomes $U(1)\times Z_2$. 
The evidence of such states comes from spontaneous magnetic fields observed in the system's bulk in muon spin relaxation experiments \cite{StanevTesanovic,carlstrom2011length, maiti2013s+,muller2018short}.
Previous theoretical studies, based on Ginzburg-Landau models, predicted such fields to arise due to certain types of defects, present in the bulk of an $s+is$ superconductor     \cite{garaud2014domain,maiti2015spontaneous,garaud2016thermoelectric, lin2016distinguishing,vadimov2018polarization,garaud2018properties,benfenati2020magnetic}.

Superconducting states that break time reversal symmetry have been sought after for decades.
Previously, the research was almost exclusively focused on different kinds of superconductors with broken time reversal symmetry (BTRS) $U(1)\times Z_2$, i.e., the topological and chiral $p+ip$ superconductors.
A hallmark of chiral superconductors that spontaneously break
time reversal symmetry are surface currents producing magnetic fields near surfaces \cite{Sigrist.Ueda:91,bouhon2010influence,etter2018spontaneous,tada2015orbital}.
By contrast, by the standard symmetry and topology arguments, $s+is$ superconductors should not have surface currents.
Namely, these are superconductors with Cooper pairing in different bands, described by several complex fields $|\Delta_\alpha|e^{\ii\theta_\alpha}$, which serve as order parameters.
The time reversal symmetry breaking is associated with a non-trivial phase difference locking between different bands $\theta_\alpha-\theta_\beta\ne 0, \pi$, so that a time reversal operation, i.e., complex conjugation of the order parameters, brings the system into a different state from which, one cannot rotate back to the original state by a gauge transformation.
The standard argument for the existence of a surface current is as follows: let us assume there is a spontaneous surface current in the superconductor. Since there is no chirality in real space in a $s+is$ state, flipping the sample does not change the chirality of the state and should not invert the current direction. Thus, one would conclude that the edge currents should be absent.

The physics of the boundary of superconductor is subtle,
and it was recently shown that there are
effects which are missed by quasi-classical approaches 
\cite{barkman2019surface,samoilenka2020bcs,samoilenka2020microscopic,benfenati2021boundary}.

In this paper, we investigate in a fully microscopic model, the physics of the boundary of
a superconductor with frustrated interband 
interaction that under certain conditions gives
rise to the $s+is$ superconducting state in the bulk.

We find superconducting surface states that can break time reversal symmetry locally near the sample's boundaries, specifically, in correspondence with the corners. 
These are counterparts of non-topological boundary states
recently reported in non-BTRS systems \cite{barkman2019surface,samoilenka2020bcs,samoilenka2020microscopic,benfenati2021boundary}.
Next we show that  non-topological, non-chiral BTRS superconducting states, such as $s+is$ states, do have spontaneous currents and spontaneous magnetic fields. These fields and currents have dipolar structures and are allowed by symmetry.
\section{The model}
\begin{figure}[htb]
    \includegraphics[width=0.5\columnwidth]{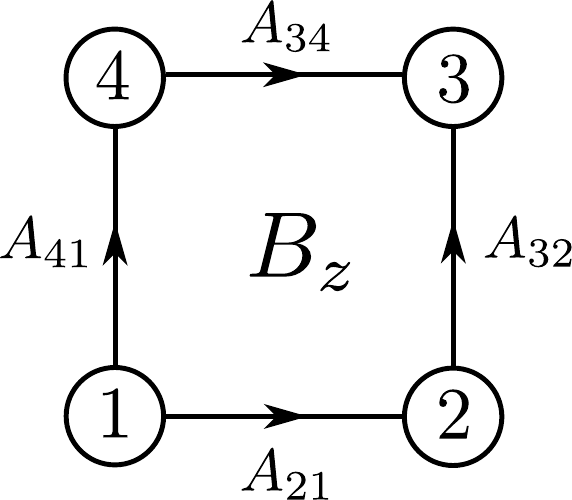}
    \caption{
    The discretized vector potential $A_{ij}$ is defined on the links connecting lattice points, whereas the magnetic field is defined on the plaquettes. They are related by the discrete curl operation, which in this case results $B_z  = A_{21} + A_{32} - A_{34} - A_{41}$.
    }
    \label{fig:latticeSetup}
\end{figure}

We obtain self-consistent solutions in a three-band Bogoliubov-de Gennes model with a gauge field, describing a three-band $s$-wave superconductor in real space. For a two-dimensional $N$-sites square lattice, the mean-field Hamiltonian we consider reads
\begin{equation}\label{eq: mfHamiltonian}
\begin{aligned}
    H= & -  \sum_{\sigma\alpha}\sum_{<ij>} \exp\qty(\ii q A_{ij} ) c^\dagger_{i \sigma \alpha } c_{j\sigma } \\
    & + \sum_{i \alpha} \qty( \Delta_{i \alpha} c^\dagger_{\uparrow i \alpha} c^\dagger_{\downarrow i \alpha} + \Delta_{i \alpha}^* c_{\downarrow i \alpha} c_{\uparrow i \alpha} )\,.
\end{aligned}
\end{equation}
The indices $i,j$ label the lattice sites, and the sum over $<i,j>$ is restricted to the nearest neighbors.  $c_{\alpha i \sigma}, c^\dagger_{\alpha i \sigma}$ are the annihilation and creation operators for a particle with spin $\sigma \in \uparrow, \downarrow$, at site $i\in [0,N-1]$ and in band $\alpha\in[1,3]$.
Moreover, the Hamiltonian is rescaled such that all energies are expressed in units of the hopping energy, which, therefore, becomes unitary, and the spatial coordinates are expressed in units of the lattice spacing. 
Finally, in this paper, the Fermi energy for each band is set to zero. 
The phase factor $\exp\qty(\ii q A_{ij})$ introduces the coupling to the vector potential, with $q$ as coupling constant, through the Peierls substitution. $A_{ij}$ is defined as 
\begin{equation}
    A_{ij} = \int_j^i \vb{A} \cdot \dd{\boldsymbol{\ell}}\,,
\end{equation}
where $\vb{A}$ is the vector potential, which is related to the magnetic field by $\curl \vb A = \vb B$.
Finally, $\Delta_{\alpha } = \abs{\Delta_{\alpha}}e^{\ii \theta_{\alpha}}$ are the superconducting gaps, which are obtained through the self-consistency equations,
\begin{equation}
(\Delta_{\alpha})_i = \sum_{\beta=1}^3V_{\alpha\beta}\langle c_{i\uparrow\beta}c_{i\downarrow\beta}\rangle \,,
\end{equation}
where $V_{\alpha\beta}=V^*_{\beta\alpha}$ is the matrix containing the intra-band ($V_{11},V_{22},V_{33}$) and inter-band couplings ($V_{12}, V_{13}, V_{23}$). 
The boundary conditions that we utilize correspond to lattice termination (see a more detailed discussion of boundary conditions in,  e.g., Refs. \cite{samoilenka2020bcs,samoilenka2020microscopic}). We neglect the effect of interband scattering at the boundary that  leads to the effects studied in Refs. \cite{bascones2001surface,Bobkov2011}.
At each iteration for $\Delta_{\alpha i}$ we compute the current density from site $j$ to $i$,
\begin{equation}
   \begin{aligned}
    J_{ij} & = - \expval{ \pdv{H}{A_{ij}}} = \\
    & = -2q \sum_{\alpha\sigma} \Im\qty[ \expval{c_{i\sigma\alpha}^\dagger c_{j \sigma \alpha}} \exp\qty(\ii q A_{ij} )]\,,
\end{aligned}
\end{equation}
which is defined, together with $A_{ij}$ on the links connecting lattice sites.
The current density is then used to re-compute the vector potential, as we outline below. 
First, we discretize the vector potential $\vb A$ using a finite difference method, where $\Delta x = \Delta y = 1$ so that it becomes identically equal to the quantities in the phase factors of \eqref{eq: mfHamiltonian}. 
Following this scheme, the magnetic field $B_z$ is defined on the lattice plaquettes as shown in \figref{fig:latticeSetup} and is related to the vector potential by a discrete curl operation. The magnetic field energy is
\begin{equation}
    E_{\textrm{mag}} = \frac{1}{2}\sum_{\rm{plaquettes}}  B_z^2\,.
\end{equation}
Then, we solve the discrete version of Maxwell equation $\curl\curl \vb A = \vb J$, namely
\begin{equation}\label{eq: Aeom}
    \pdv{E_{\textrm{mag}}}{A_{ij}} + \expval{ \pdv{H}{A_{ij}}} =0\,.
\end{equation}
In the computation of the vector, potential we utilize the  boundary conditions which set the magnetic field on the sample's edges equal to zero, maintaining gauge invariance.

We solve self-consistently for the gaps and the vector potential. To compute $\Delta_{\alpha i}$ we use Chebyshev polynomial expansion method \cite{kernelPoly,chebCovaci,chebNagai} with a polynomial up to order 700, which is sufficient in the considered temperature range. To calculate the vector potential at each iteration, we perform a gradient descent step, adapting $\vb A$ to the changing current density distribution. Then, we use the new vector potential to update all the phase factors in \eqref{eq: mfHamiltonian}. 
We iterate this fully self-consistent procedure until a specific convergence criterion is achieved. In our paper, we consider a simulation converged when the mean relative variations of the individual gaps and the vector potential components are smaller or equal to $10^{-8}$.

We consider two-dimensional square lattice samples of size $N_x N_y = 100\times100$, and unless specified otherwise, we fix the temperature to $T=0.44$ and and the coupling constant to the vector potential to $q=-0.5$.
The intra- and interband interactions are given by the coupling matrix,
\begin{equation}\label{eq:couplingMatrix}
    V_{\alpha\beta}=\mqty(1.92 && -1.0 && -1.0 \\ -1.0 && 1.95 && -1.0 \\ -1.0 && -1.0 && 1.9)_{\alpha\beta}\, .
\end{equation}
In $s+is$ superconductors, the interband couplings yield phase frustration. To fully minimize the energy, a phase difference of $\pi$ between each band would be preferred, which is not achievable in the three bands' case. 
The $s+is$ state arises where the disparity of the coupling is not too significant so that there are two energetically equivalent interband phase-difference locking $\theta_\alpha-\theta_\beta \neq 0, \pi$. Whereas, when the phase differences are $\pi,0$ the system is in the so-called $s_{+-}$ or  $s_{++}$ states, respectively.
The choice of relatively strong coupling is motivated by the fact that we perform a fully self-consistent simulation of a two-dimensional system.   Lager characteristic lengths, arising for inhomogeneous solutions at weak coupling, require computationally inaccessible systems sizes.
\section{The results}
\begin{figure}[htb]
    \includegraphics[width=0.99\columnwidth]{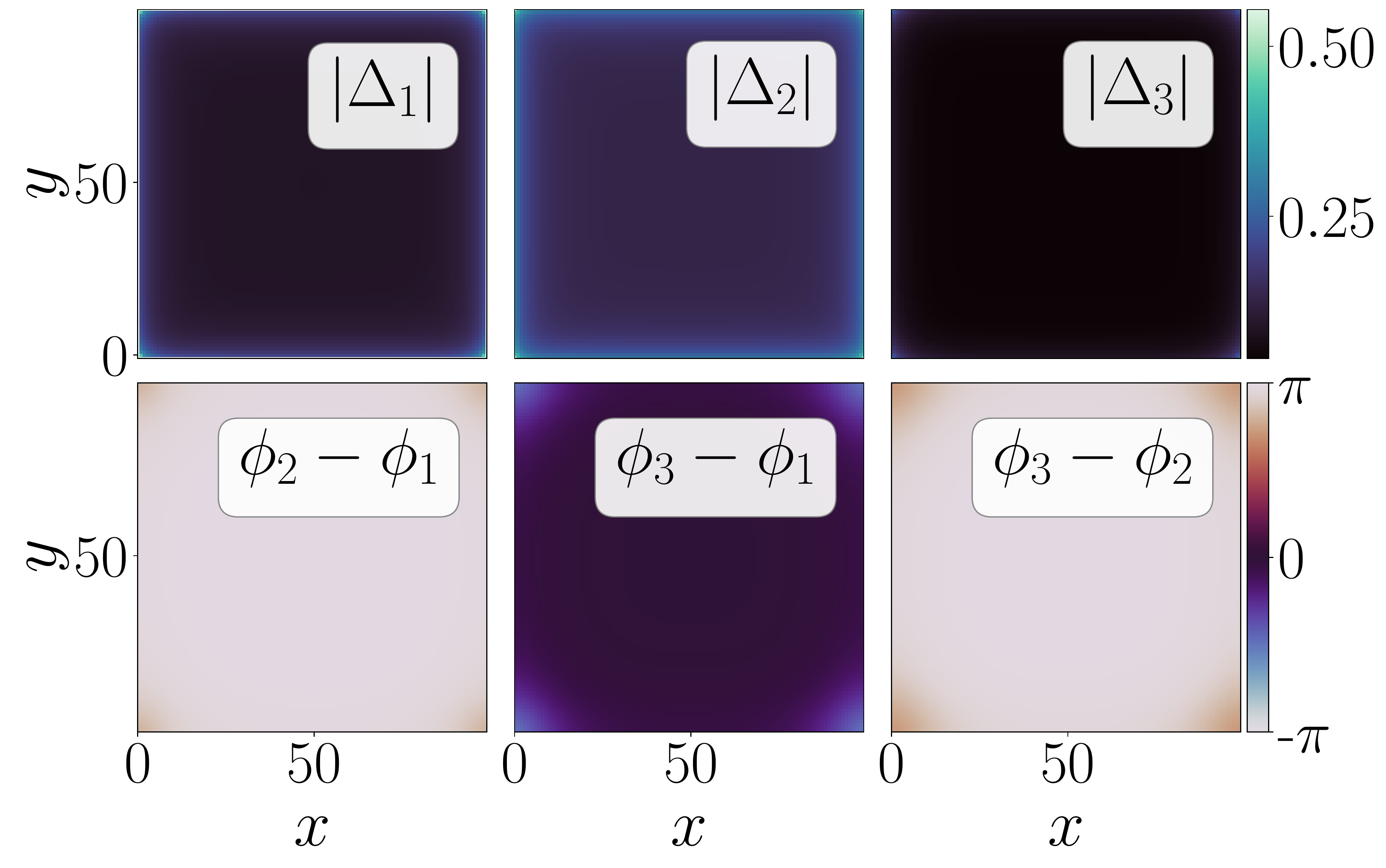}
    \caption{Superconducting gaps and phase differences for a three-band  two-dimensional superconductor with time-reversal symmetry breaking.
    We observe that the interband phase differences are not spatially uniform: whereas in the bulk the phase differences are $\pi,0$, resulting in a $s_{+-}$ state, near the boundary there is a local time reversal symmetry breaking resulting in a local $s+is$ state. This situation yields spontaneous counter-currents in different bands, arising near the corners. The parameters used in the simulation are $T=0.44$ and $q=-0.5$ and \eqref{eq:couplingMatrix} for what concerns inter/intra-band coupling.
    \label{fig:gap_square}}
\end{figure}

\begin{figure}[h]
    \includegraphics[width=0.95\columnwidth]{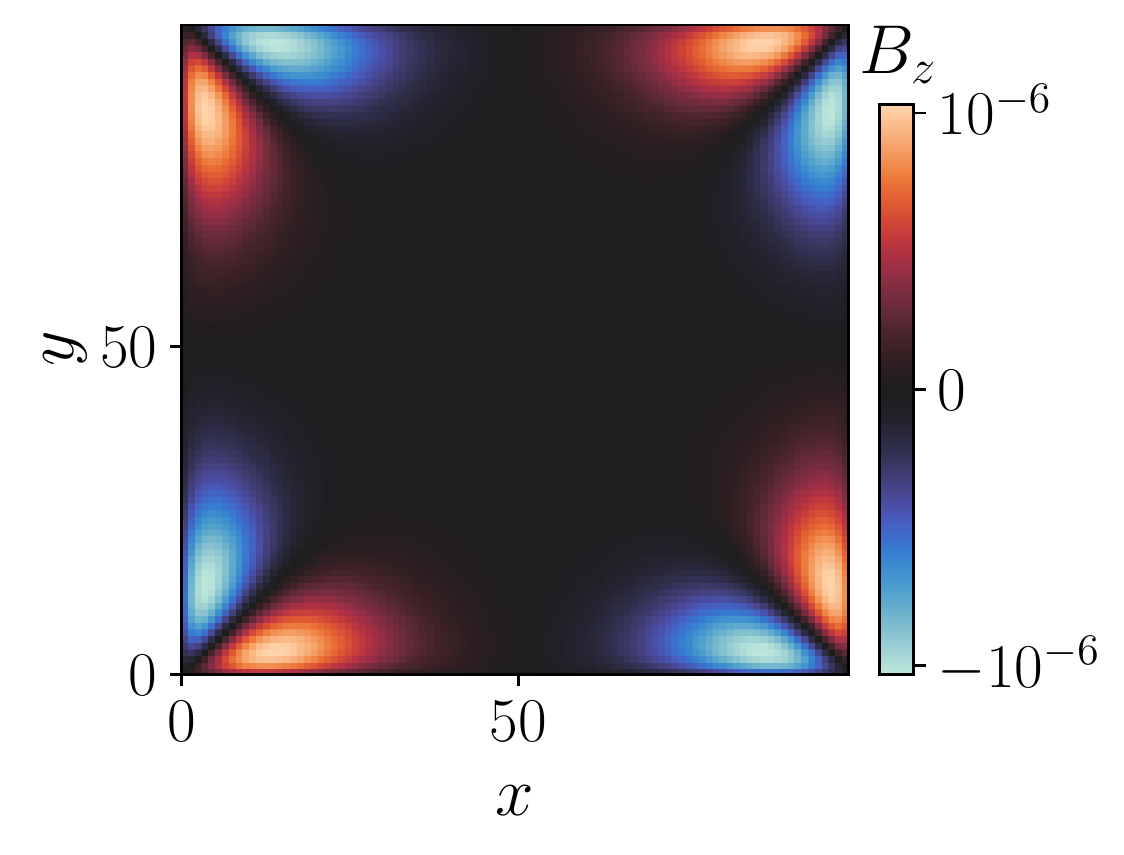}
    \caption{A spontaneous magnetic field in the corners of a square superconducting sample with local $s+is$ state, obtained via a self-consistent solution of Bogoliubov-de Gennes model. 
    The value of the magnetic flux associated to each red (positive) \textit{petal} of magnetic field is $\Phi/\Phi_0=3.4\cdot10^{-5}$, whereas the blue (negative) \textit{petals} have opposite flux.
    The parameters used in the simulation are $T=0.44$ and $q=-0.5$ and \eqref{eq:couplingMatrix} for what concerns inter/intraband coupling. The spatial extent of the magnetic flux is macroscopic as it is determined by coherence and magnetic-field penetration lengths.}
    \label{fig:B_square}
\end{figure}

\begin{figure}[h]
    \includegraphics[width=0.925\columnwidth]{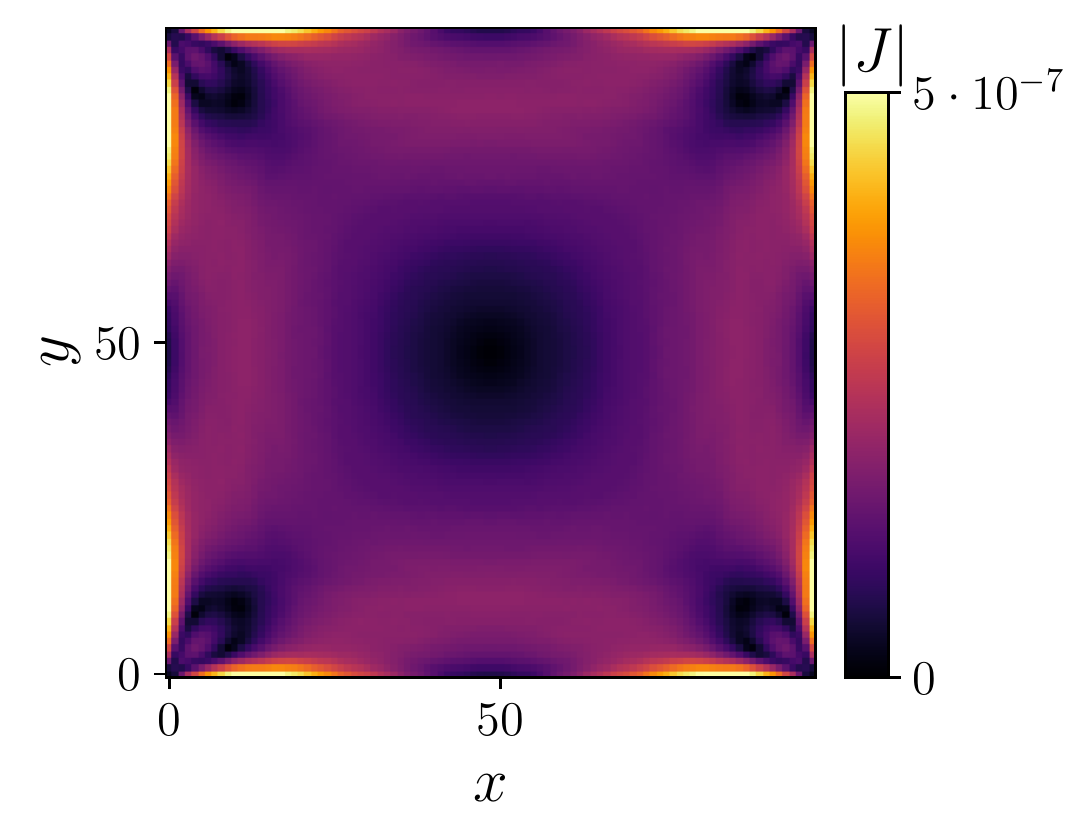}
    \caption{
    Modulus of the spontaneous supercurrent generating the magnetic field in \figref{fig:B_square}. We can notice a substantial localization of the currents near the sample's corners.
    The parameters used in the simulation are $T=0.44$ and $q=-0.5$ and \eqref{eq:couplingMatrix} for what concerns inter/intra-band coupling.
    }
    \label{fig:J_square}
\end{figure}

We begin by analyzing the case of a square sample.
First, we find that, at elevated 
temperature, the system has a new phase, in addition to the phases discussed for the same model in  Refs. \cite{StanevTesanovic,carlstrom2011length,maiti2013s+}.
In that state, the bulk of the system is in a $s_{+-}$ state, but the system breaks time reversal symmetry, locally, near the boundaries, where $\theta_\alpha-\theta_\beta \neq 0, \pi$.
The resulting gaps absolute values and phase differences are shown in \figref{fig:gap_square}. 
Second, our solutions show the presence of a spontaneous magnetic field, localized near sample corners shown in \figref{fig:B_square}.
The modulus of the supercurrent generating the magnetic field is displayed in \figref{fig:J_square}.
Both  effects originate from the simultaneous enhancement of the density of states near the sample's boundaries \cite{samoilenka2020bcs,benfenati2021boundary}, and by the $s+is$ state localized at the corners.
We find that the effect of spontaneous field in our model is not generic but exists 
when there is a slight disparity in the couplings of different bands. 
In each corner, the spontaneous magnetic field has a dipolar structure, and, therefore, it carries zero net flux through the whole system. 
The field configuration respects the rotation symmetry of the lattice.
The origin of the spontaneous magnetic field can be understood as follows.
First, in an $s+is$ superconductor, the normal modes are a mixed linear combination of the gap amplitudes and phase differences \cite{carlstrom2011length}, which means that even a tiny variation of relative densities
results in a variation of relative phases. This is in contrast with ordinary multiband superconductors where minor spatial variations of the gap amplitudes do not produce variations in the phase difference.

In the present system, the gradients of relative densities and relative phases induce supercurrents.
Let us consider, for example, a Ginzburg-Landau model for a two-dimensional three-band superconductor.
The expression for magnetic field $B_z$ can
be written by taking the curl of the vector potential, expressed as a function of the supercurrent. For a three-band superconductor with standard gradient terms, the expression reads \cite{garaud2013chiral}
 \begin{equation}\label{eq:BGL}
 \begin{split}
     B_z = -\varepsilon_{ij}\partial_i\qty(\frac{J_j}{e\abs{\Psi}^2})- \frac{\ii\varepsilon_{ij}}{e\abs{\Psi}^4}\left(\abs{\Psi}^2\partial_i\Psi^\dagger\partial_j\Psi\right. \\
   + \Psi^\dagger\partial_i\Psi\partial_j\Psi^\dagger\Psi\Big)\,.
 \end{split}
 \end{equation}
This is a three component generalization of the results in \cite{Babaev.Faddeev.ea:02}, where $\Psi=(\psi_1,\psi_2,\psi_3)$ is a vector with the three order parameters as components and modulus $\abs{\Psi}^2 = \Psi^\dagger \Psi$. $J_i$ is the $i$-th spatial component of the Ginzburg Landau current density and $\varepsilon_{ij}$ is the two dimensional version of the Levi-Civita symbol. 
The first term in \eqref{eq:BGL} is the standard contribution, generic for London's  magneto statics. The second term is specific for three-band superconductors and describes currents originating from the cross gradients of the relative phases and relative amplitudes of the gaps in different bands.
It has the form of $\mathbb{C}P^2$ skyrmionic topological charge density \cite{garaud2013chiral},
\begin{equation}\label{eq:topQ}
    \mathcal{Q}\qty(\Psi) = \int_{\mathbb{R}^2}\frac{\ii\varepsilon_{ij}}{2\pi\abs{\Psi}^4}\left(\abs{\Psi}^2\partial_i\Psi^\dagger\partial_j\Psi\right. \\
   + \Psi^\dagger\partial_i\Psi\partial_j\Psi^\dagger\Psi\Big)\dd^2 x \, .
\end{equation}
Note that the second term is identically zero if there is no disparity in the variations of the gaps in different bands.

Near the surfaces and corners, Friedel oscillations of the density of states produce disparities in the gap amplitudes of the different bands \cite{samoilenka2020bcs,benfenati2021boundary}. However, when the gradients of these quantities are collinear, the second term in \eqref{eq:BGL} remains zero and, thus, one does not see any currents in the vicinity of the edges.
Our microscopic solutions show that, in the corners, the spatial profile of the gaps exhibits noncollinear gradients in the amplitude and phase difference and, therefore,  generates spontaneous currents. 
The gap enhancement was demonstrated also to arise at various boundaries in higher dimensions in single-component systems \cite{samoilenka2020bcs}. We, therefore, expect a similar effect to be present near the edges and vertices of a three-dimensional superconductor.

\begin{figure}[htb]
    \includegraphics[width=0.90\columnwidth]{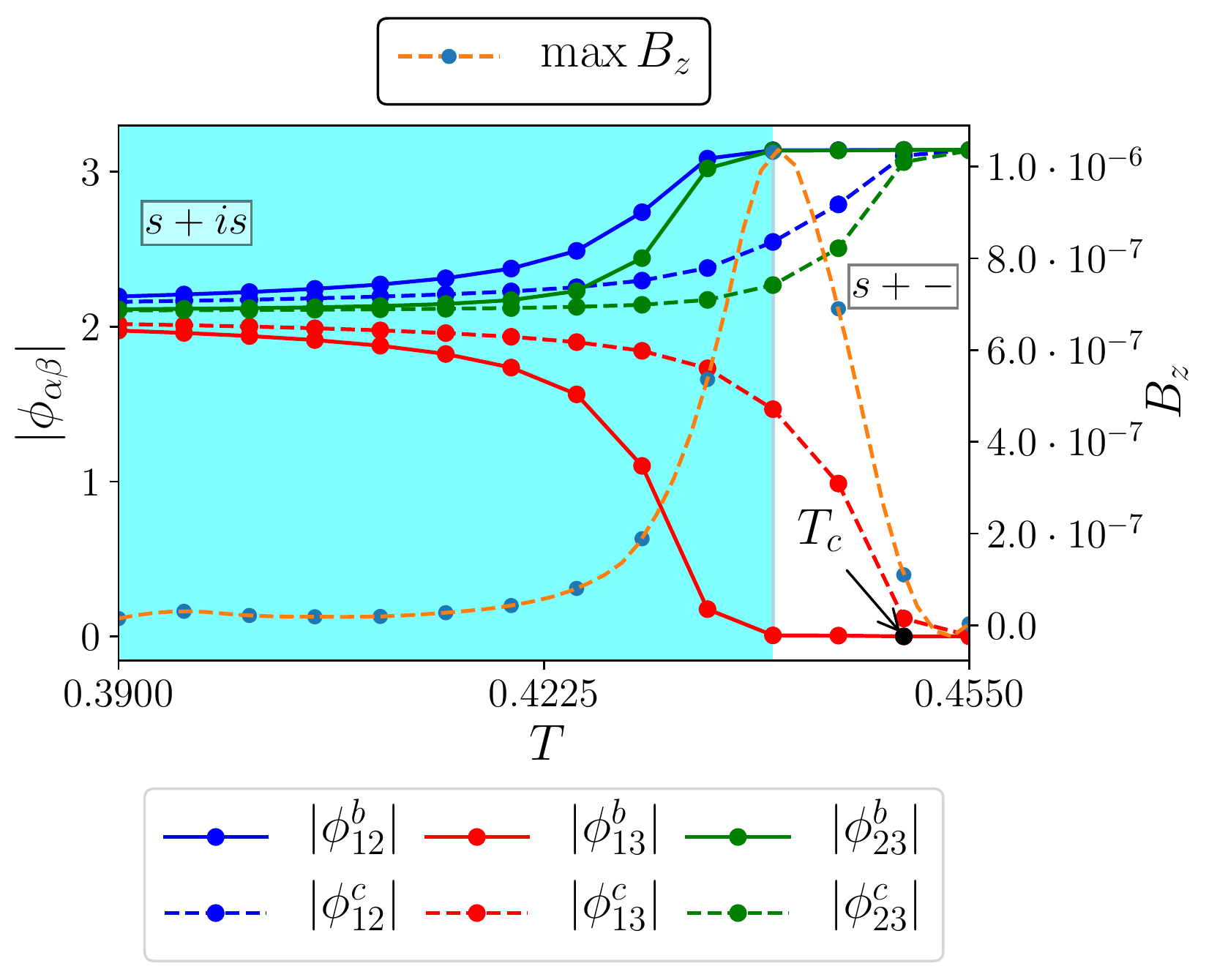}
    \caption{
    Temperature dependence of the absolute values of the phase differences $|\phi_{\alpha \beta}| = |\phi_{\alpha} - \phi_{\beta}|$, with $\alpha,\beta=1-3$ in the system's bulk $|\phi_{\alpha \beta}^\textrm{b}|$ and in the system's corners $|\phi_{\alpha \beta}^\textrm{c}|$. 
    The figure also reports the maximal spontaneous field $B_z$, which exhibits a non-monotonic behavior as a function of the temperature $T$.
    The effect is the strongest at the transition point in which, the system's bulk turns from $s+is$ (blue background) to $s_{+-}$ (white background); at this point, time reversal symmetry is locally broken in the vicinity of the corners.
    Furthermore, we observe that the TRSB state persists in the corners even after it disappears from the bulk.
    This result illustrates that spontaneous magnetic signatures significantly depend on the system parameters and, in this example, are best detectable at a higher temperature.
    It is important to note that this effect occurs below the system's bulk critical temperature $T_c$.
    The parameters used in the computations are $q=-0.5$ and \eqref{eq:couplingMatrix} for what concerns inter/intraband coupling.}
    \label{fig:B_T}
\end{figure}

Let us consider now how the spontaneous magnetic field $B_z$ varies as the temperature $T$ changes. Figure \ref{fig:B_T} displays the maximum value of $B_z$ for temperatures in the range of $T\in[0.39, 0.455]$ in a square superconducting sample.
Figure \ref{fig:B_T} also depicts the absolute values of the phase differences $|\phi_{\alpha \beta}| = |\phi_{\alpha} - \phi_{\beta}|$, with $\alpha,\beta=1-3$, in the bulk ($|\phi_{\alpha \beta}^\textrm{b}|$) and in the corners ($|\phi_{\alpha \beta}^\textrm{c}|$). 
The time reversal symmetry breaking superconductivity survives at slightly higher temperature than the bulk critical temperature in the vicinity of the corners.
\begin{figure*}[htb]
    \includegraphics[width=1.99\columnwidth]{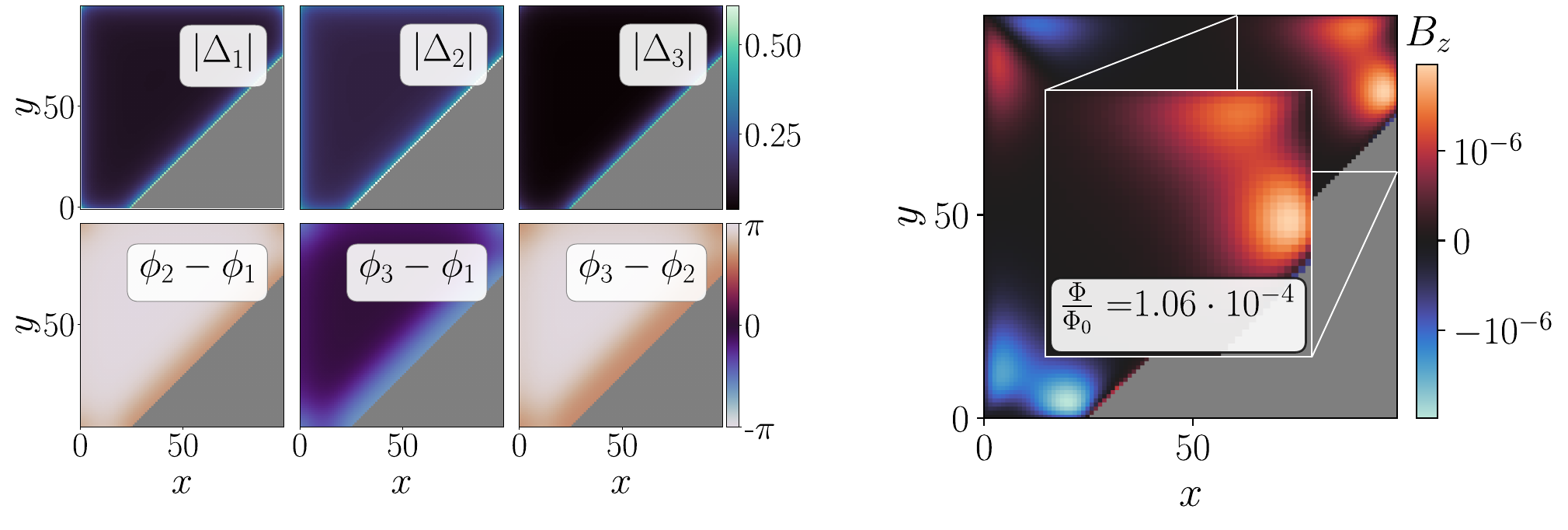}
    \caption{Superconducting gaps and phase differences (on the left) and spontaneous magnetic field (on the right) for a three-band two-dimensional superconductor with time-reversal symmetry breaking. Differently from \figref{fig:gap_square}, the sample presents a diagonal cut along which the three order parameters undergo enhancement compared to the horizontal sides. The gray color indicates the vacuum. 
    We can notice that in the corners formed with the diagonal edge, the magnetic field is enhanced and does not have a locally dipolar profile, in contrast to the squared sample of \figref{fig:B_square}.
    The parameters used in the simulation are  $T=0.44$ and $q=-0.5$ and \eqref{eq:couplingMatrix} for what concerns inter/intraband coupling.}
    \label{fig:gaps_cut}
\end{figure*}
The result in \figref{fig:B_T} suggests that the presence of spontaneous magnetic signatures is not a universally detectable property of three-band superconductors breaking time reversal symmetry, and it
may be easier to detect at elevated temperatures.
However, it is peculiar to note that the maximal spontaneous magnetic signature occurs at the same temperature at which the bulk is no longer in the $s+is$ state.

For a square geometry, the flux in each corner locally adds up to zero, which may compromise the detection process due to the resolution in scanning superconducting quantum interference device (SQUID) probes.
To make the effect more observable, one may break the spatial symmetry by considering different shapes.

By cleaving a corner of a square, one obtains the geometry with five corners shown in \figref{fig:gaps_cut}, where the gray color indicates the vacuum. 
In which case, the total flux is still zero. However, the corner states inherently depend on the corner geometry, and the resulting flux fractionalization pattern becomes different: now there are well-separated corners with nonzero local flux. 
In this configuration, we notice how the magnetic field (left panel) maintains the same spatial profile of \figref{fig:B_square} in the 90\textdegree corner but substantially changes near the diagonal edge. The magnetic flux in the lower left and upper right corners of the sample is on the order of $\sim 10^{-4}$ flux quanta and does not have a locally dipolar structure. We expect such magnetic fields to be more easily detectable by SQUIDS techniques.   
\section{Conclusions}
In conclusion,
we considered boundary effects in the basic microscopic model of a three-band superconductor with repulsive interband interaction.
We find that, in the minimal model, that does not include interband impurity scattering at the boundaries, the critical temperature of the time reversal symmetry breakdown is different for the boundaries of 
a superconductor and its bulk due to the presence of boundary states.
An interesting followup investigation would combine the effect of the interband surface scattering, considered at the level of a quasiclassical 
theory in Ref. \cite{Bobkov2011}, with the boundary effects considered in our paper, that appear in a fully microscopic theory beyond the usual quasiclassical approximation.

Our second finding is that $s+is$ superconducting states localized near the sample's boundaries for certain parameters give rise to spontaneous
boundary currents. 
This occurs at lower-dimensional boundaries: near the corners in the two-dimensional case, and we expect in the vicinity of the edges of three-dimensional samples. The origin of these fields is the existence of surface states \cite{samoilenka2020bcs,samoilenka2020pair,benfenati2021boundary,samoilenka2020microscopic} and the mixing of gap amplitude and phase-difference modes \cite{carlstrom2011length} occurring in superconductors breaking time reversal symmetry. 
This phenomenon is different in its origin and form from the surface currents in topological chiral superconductors \cite{Sigrist.Ueda:91,bouhon2010influence,etter2018spontaneous,tada2015orbital}
.

In our example, we find that, the spontaneous fields are sufficiently strong and can be detected  by
scanning SQUID techniques \cite{hicks2010limits}, scanning Hall \cite{gutierrez2012scanning}, or single-atom magnetic resonance \cite{Willke2019}.
Since the spontaneous fields originate from the interband phase-difference gradients, they are expected to persist and, thus, to serve as a probe of
the $Z_2$ bosonic metal phase \cite{bojesen2013time,bojesen2014phase}, that was recently reported in  Ba$_{1-x}$K$_x$Fe$_2$As$_2$ at $x\approx 0.8$ \cite{grinenko2021bosonic}.

This superconducting state is nonchiral and nontopological within the common classification framework.
That shows that the boundary currents is a more general phenomenon that can exist in nontopological systems.
We note, however, that our simulations show that the effect exists and is detectable, only within certain parameter rages and, therefore, is not generic.

\section{Acknowledgements}
We thank M. Barkman and A. Samoilenka for discussions and collaboration. The work was supported by the Swedish Research Council Grants No. 2016-06122, 2018-03659, and No. the G\"{o}ran Gustafsson Foundation for Research in Natural Sciences and Medicine.
\bibliography{bibliography.bib}
\end{document}